\documentclass[preprint,3p, times]{elsarticle}

\usepackage{amssymb, bm}
\usepackage[figuresright]{rotating}
\usepackage{graphicx}
\usepackage{epsf}
\usepackage{times}
\usepackage{float}
\usepackage{amsmath}
\usepackage{rotating}
\usepackage{color}
\usepackage[ruled, vlined]{algorithm2e}
\usepackage{tikz}
\usetikzlibrary{shapes, snakes, positioning,fit,calc,math,patterns}

\def\defeq{\stackrel{\triangle}{=}}
\newcommand{\boldgk}[1]{\mbox{\boldmath $#1$}}

\newcommand{\expd}[1]{e^{\displaystyle #1}}

\graphicspath{figures/}

\begin{document}

\begin{frontmatter}

\title{An Anchor-Point Based Image-Model for Room Impulse Response Simulation with Directional Source Radiation and Sensor Directivity Patterns}


\author[ciaic,marineschool]{Chao Pan}
\author[huawei]{Lei Zhang}
\author[ciaic,marineschool]{Yilong Lu}
\author[ciaic,marineschool]{Jilu Jin}
\author[huawei]{Lin Qiu}
\author[ciaic]{Jingdong Chen}
\author[inrs]{Jacob Benesty}
\address[ciaic]{Center of Intelligent Acoustics and Immersive Communications, Northwestern Polytechnical
   University, China}
\address[marineschool]{School of Marine Science and Technology, Northwestern Polytechnical
   University, China}
\address[huawei]{Huawei Xiliubeipo Village, Xiliubeipo Being Village, Hiking Road, High-tech Industrial Development Zone, Dongguan, Guangdong 523808, China}
\address[inrs]{INRS-EMT, University of Quebec, Canada}

\begin{abstract}
The image model method has been widely used to simulate room impulse responses and the endeavor to adapt this method to different applications has also piqued great interest over the last few decades. This paper attempts to extend the image model method and develops an anchor-point-image-model (APIM) approach as a solution for simulating impulse responses by including both the source radiation and sensor directivity patterns.  To determine the orientations of all the virtual sources, anchor points are introduced to real sources, which subsequently lead to the determination of the orientations of the virtual sources. An algorithm is developed to generate room impulse responses with APIM by taking into account the directional pattern functions, factional time delays, as well as the computational complexity. The developed model and algorithms can be used in various acoustic problems to simulate room acoustics and improve and evaluate processing algorithms.
\end{abstract}


\end{frontmatter}

\section{Introduction}%

{R}{oom} acoustics simulation plays an important role in many applications on development and evaluation of such algorithms as acoustic source localization, noise reduction, speech enhancement, source separation, dereverberation, speech data augmentation for machine learning and artificial intelligence, to name but a few.
It has been intensively studied for decades \cite{Allen1979,Savioja2015,Vaelimaeki2012,savioja1999creating}. Among the different methods that have been developed, the so called image model method \cite{Allen1979}, which is sometimes also called the image method, the image-source method, or simply the image model, has been dominantly used to simulate the impulse response from a given source to a sensor position. Many efforts have also been devoted to generalizing this useful technique. For example, the work in \cite{Borish1984} extended the method to deal with arbitrary polyhedra room geometries. In  \cite{Lehmann2008,lehmann2009diffuse}, the energy decay behavior was investigated, and a fast algorithm was developed to generate single-channel impulse responses. In \cite{jarrett2012rigid}, impulse responses for rigid spherical microphone arrays were derived. Moreover, the problem of how to achieve fractional time delay has been studied to simulate the impulse responses for microphone array processing to better model the reflection paths \cite{peterson1986simulating,savioja1999creating, betlehem2012sound,kompis1993simulating, samarasinghe2018spherical}.
In the work presented in \cite{brinkmann2019extending}, an extension of the image model was introduced to address the problem of room impulse response simulation for sources with directional radiation patterns, which is achieved by including the formulation of exit angles for individual image sources. However, the practical implementation of this method requires a reformulation of both source radiation and sensor directivity patterns every time when their orientations are changed. Furthermore, the validity of the source exit angles outlined in \cite{brinkmann2019extending} is limited to scenarios where the room geometry is rectangular.

This paper attempts to improve and enhance the image model method by taking into consideration of both the source radiation pattern and the sensor directivity pattern, An approach called the anchor-point-image-model (APIM) is developed\footnote{Detailed description and the Matlab code can be downloaded from https://github.com/ChaoTR2GH/AnchorPointImageModel.git}, which manipulates the coordinate systems of every image sources to achieve highly efficient calculations of source orientations. Specifically, to ascertain the coordinate systems of the image sources and subsequently compute their orientations, the so-called anchor points are introduced to real sources, which subsequently lead to the determination of the orientations of the virtual sources. An algorithm is then developed to generate room impulse responses in the framework of APIM by taking into account the sensor's directivity, sub-sample or fractional delays, pattern function modeling, and computational complexity. In comparison with the existing methods, the APIM and the presented algorithm exhibits more flexibility from the following perspectives.
\begin{itemize}
\item	\emph{Simple specification:} Only the pattern functions of the sources and sensors need to be specified. The orientation adjustment is automatically achieved through configuring anchor points.

\item	 \emph{Image model framework:} The developed model and algorithm are similar to the original image model method, which can be easily followed by the reader who is familiar with the classical image model method.

\item	 \emph{Intuitive implementation:} The orientation of all the real and image sources are readily comprehensible, which enables intuitive understanding of the contribution of every image source to the sensor observation.

\item	 \emph{Wide-range applicability:} The APIM approach take into consideration of both the source radiation pattern and the sensor directivity pattern as well as the orientation of those patterns. It is flexible to simulate any type of room impulse responses.   	

\item    \emph{Fractional Delays:} Fractional delays are considered in the developed algorithm so it can be use simulate room acoustics for microphone array processing.

\item	 \emph{General room geometry:} The developed method is not limited to the rectangular room geometries and, can be used in any room geometry as long as the image source positions can be computed.
 
\end{itemize}
 
The remainder of the paper is organized as follows. In Section~\ref{sect-SM}, we present the framework of the image model method, and describe the impact of the source radiation pattern and sensor directivity pattern. In Section~\ref{sect-Coordiante-ort}, we present the coordinate systems for both the sensor and the real/virtual sources, the definition of the orientation vector and angles, and a method to determine the orientation of source and sensors by introducing some anchor points. In Section~\ref{sect-alg-implemt}, the implementation of the proposed APIM approach is described, which includes the radiation/directivity pattern model, the fractional time delay, and the detailed algorithm implementation. In Section~\ref{sect-examples}, several examples are presented to show how the radiation/directivity pattern may affect the impulse response from the source to the sensor. Finally, we important conclusions are given in Section~\ref{sect-cnls}.

\section{ Image Model Method and Problem Illustration}
\label{sect-SM}

The core idea of the image model method is to treat the reflections of the source as virtual sources (also called images) around the sensor. Consider a room of size $L_{x}\times L_{y} \times L_{z}$, where the 3-dimensional (3D) Cartesian coordinate system is used to express positions. The position of the microphone is denoted as
\begin{align} 	
\mathbf{r}_{\mathrm{mic}}
	\defeq  \left[\begin{array}{ccc}
		x_{\mathrm{mic}}& y_{\mathrm{mic}}& z_{\mathrm{mic}}
	\end{array} \right]^{T},
\end{align}
where the subscript $_\mathrm{mic}$ stands for microphone sensor and the superscript $^T$ is the transpose operator.
Assume that a source is located at the position:
\begin{align}
	\mathbf{r}_{\mathrm{s}}
	\defeq   \left[\begin{array}{ccc}
		x_{\mathrm{s}}& y_{\mathrm{s}}& z_{\mathrm{s}}
	\end{array} \right]^{T},
\end{align}
where the subscript $_\mathrm{s}$ stands for source. According to \cite{Allen1979,Huang2006}, the position of the images are
\begin{align}
	\mathbf{r}_{q_{x},q_{y},q_{z}}^{p_{x}, p_{y}, p_{z}}
	=  \left[\begin{array}{c}
		(-1)^{p_{x}} x_{\mathrm{s}} + 2 q_{x} L_{x}\\
		(-1)^{p_{y}} y_{\mathrm{s}} + 2 q_{y} L_{y}\\
		(-1)^{p_{z}} z_{\mathrm{s}} + 2 q_{z} L_{z}
	\end{array}\right], \label{def-r-pq}
\end{align}
where $q_{x}=-Q_{x},\ldots,Q_{x}$, $q_{y} = -Q_{y},\ldots,Q_{y}$, $q_{z} = -Q_{z},\ldots,Q_{z}$, $p_{x}=0,1$, $p_{y}=0,1$, $p_{z}=0,1$, and $Q_{x}$, $Q_{y}$, and $Q_{z}$ are three positive integers. Let us further define the following set:
\begin{align}
	\mathbb{I}
	\defeq \left\{(p_{x}, p_{y}, p_{z}; q_{x}, q_{y}, q_{z}), \forall p_{x}, p_{y}, p_{z}, q_{x}, q_{y}, q_{z} \right\}.
\end{align}
If we denote the total number of elements in $\mathbb{I}$ as $N$, it can be checked that $N=8(2Q_{x}+1)(2Q_{y}+1)(2Q_{z}+1)$. Now, let us denote the position of the $n$th image (the $n$th element of $\mathbb{I}$) as
\begin{align}
	\mathbf{r}_{n}
	= \mathbf{r}_{q_{x},q_{y},q_{z}}^{p_{x}, p_{y}, p_{z}}, \ n=0,1,\ldots, N-1. \label{def-rn}
\end{align}
By assuming that the source and sensor are omnidirectional, the traditional image model method generates the impulse response according to
\begin{align}
  \label{image-model-IR}
	h \left( t \right)
	= \sum_{n=0}^{N-1}  	\dfrac{1}{4\pi d_{n}} \beta_{n}
    \delta\left(t-\tau_{n}\right),
\end{align}
where
\begin{align}
	\tau_{n}
	&\defeq  \dfrac{d_{n}}{c},\\
	d_{n}
	&\defeq  \left\|\mathbf{r}_{n} - \mathbf{r}_{\mathrm{mic}} \right\|, \label{def-dn}\\
	\beta_{n}
	&\defeq  \beta_{x_{0}}^{|q_{x}-p_{x}|} \beta_{x_{1}}^{|q_{x}|}
	\beta_{y_{0}}^{|q_{y}-p_{y}|} \beta_{y_{1}}^{|q_{y}|}
	\beta_{z_{0}}^{|q_{z}-p_{z}|} \beta_{z_{1}}^{|q_{z}|}, \label{def-beta-n}
\end{align}
with $\left\| \cdot \right\|$ being the Euclidean distance. The value of $\tau_{n}$ is the propagation time from the position $\mathbf{r}_{n}$ to the sensor position $\mathbf{r}_{\mathrm{mic}}$, $c$ is the speed of sound in the air, $d_{n}$ is the distance from the $n$th image source to the microphone, and $\beta_{x_{0}}, \beta_{x_{1}}, \beta_{y_{0}}, \beta_{y_{1}}, \beta_{z_{0}}$, and $\beta_{z_{1}}$ are the reflection coefficients (or attenuation factors) of the six walls.

However, in most practical applications, the sources and sensors may not be omnidirectional. The sources may radiate sounds with a certain directivity pattern, and the sensor can be directional as well. To incorporate the source radiation pattern and the sensor directivity pattern into the image model method, let us transform the impulse response in  (\ref{image-model-IR}) into the frequency domain, thereby giving the following transfer function:
 \begin{align}
 \label{image-model-tf}
 	H \left( f \right)
 	 = \sum_{n=0}^{N-1}
	\dfrac{1}{4\pi d_{n}} \beta_{n}
	\expd{-\jmath 2\pi f \tau_{n}},
 \end{align}
where $f$ is the temporal frequency in hertz (Hz) and $\jmath$ is the imaginary unit. Considering the source radiation pattern as well as the sensor directivity pattern, one can extend (\ref{image-model-tf}) to a more general form as
 \begin{align}
 \label{image-moodel-TF}
 	H \left( f \right)
 	 =  \sum_{n=0}^{N-1}
	\dfrac{1}{4\pi d_n} \beta_n \mathcal{A}_{n}(f) \mathcal{B}_n \left( f \right)
	\expd{-\jmath 2\pi f \tau_n},
 \end{align}
where $\mathcal{A}_{n}(f)$ and $\mathcal{B}_{n}(f)$ are the filters modeled by the directional patterns of the sensor and source, respectively, which are determined by the radiation/directivity patterns and the relative positions between the real/virtual sources and the sensor. For convenience, we call   $\mathcal{A}_{n}(f)$ and $\mathcal{B}_{n}(f)$ as pattern responses of the sensor and source, respectively.

\begin{figure}[t]
    \centering
	\includegraphics{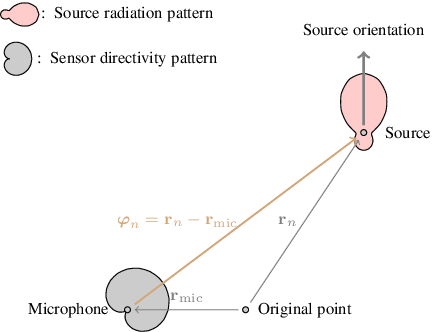}\\
	\caption{Illustration of the core parameter $\boldgk{\varphi}_{n}$ and the orientations of the source and sensor, where $\mathbf{r}_{n}$ is the position of the $n$th source and $\mathbf{r}_{\mathrm{mic}}$ is the position of the microphone.} \label{fig-illust-coreparamter}
\end{figure}

Before leaving this section, let us define a direction vector of the $n$th image relative to the sensor as
\begin{align}
	\boldgk{\varphi}_{n}
	\defeq \mathbf{r}_{n} - \mathbf{r}_{\mathrm{mic}}.
	\label{def-varphi-n}
\end{align}
The distance from the $n$th image to the sensor is $d_n = \left\| \boldgk{\varphi}_{n} \right\|$.
The relationship between $\boldgk{\varphi}_{n}$, $\mathbf{r}_{n}$, and $\mathbf{r}_{\mathrm{mic}}$ is illustrated in
Fig.~\ref{fig-illust-coreparamter}. It should be noted that $\boldgk{\varphi}_{n}$ is the core parameter to determine $\mathcal{A}_{n}(f)$ and $\mathcal{B}_{n}(f)$ in the proposed approach.

\section{Coordinate Systems and Orientations}
\label{sect-Coordiante-ort}

\begin{figure}[t]
	\centering
	\includegraphics{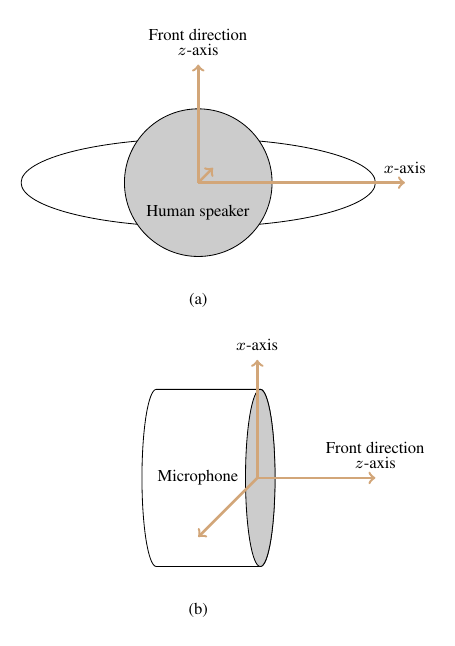}
	\caption{Illustration of the coordinate system for a human speaker (view from the top, and the front direction being the direction towards which the speaker talks) and the sensor: (a) human speaker and (b) sensor.}
	\label{fig-front-direction}
\end{figure}

Without loss of generality, we assume that the radiation or the directivity pattern of the source/sensor is oriented in such a way that the main lobe (called the front direction, see Fig. \ref{fig-illust-coreparamter}) coincides with the $z$-axis; in this case, the elevation angle $\theta$ is the angle between the front direction and the measured direction $(\theta, \phi)$. Then, once the $x$-axis is specified, the coordinate system is determined. We present in the following some examples how to choose the $x$-axis.
 \begin{itemize}
 	\item For a human speaker, the $x$-axis is chosen as the line parallel to the shoulders, as shown in Fig.~\ref{fig-front-direction}(a).
 	\item For a sensor, the direction of the $x$-axis depends on how this sensor is mounted on the device. One way is to choose a line in the sensor surface plane, which is orthogonal to the $z$-axis, as the $x$-axis, as shown in Fig.~\ref{fig-front-direction}(b).
 \end{itemize}
 To simplify the implementation, the directivity pattern of the sensor is assumed symmetric with respect to the $z$-axis.
Now, let us denote the $x$-axis, $y$-axis, and $z$-axis directions as $\bm{i}$, $\bm{j}$, and $\bm{k}$, respectively. The $x$-axis of the $n$th source is denoted as $\bm{i}_{n}$ and the $x$-axis of the sensor is denoted as $\bm{i}_{\mathrm{mic}}$. In case that $x$-axis (i.e., $\bm{i}$) and $z$-axis (i.e., $\bm{k}$) are  determined, the $y$-axis (i.e., $\bm{j}$) can be achieved by rotating the $x$-axis $90^{\circ}$ around the $z$-axis. According to the Rodrigues formula, we have
 \begin{align}
 	\bm{j}
 	&= \bm{i} \cos 90^{\circ} + \mathbf{T}_{\bm{k}} \bm{i}\sin 90^{\circ} + \bm{k}\left(\bm{k}^{T}\bm{i}\right)\left(1-\cos 90^{\circ}\right)\\
 	&= \mathbf{T}_{\bm{k}} \bm{i} + \bm{k}\left(\bm{k}^{T}\bm{i} \right)\\ 
 	&= \mathbf{T}_{\bm{k}} \bm{i}, \label{y-axis-gen}
 \end{align}
 where (\ref{y-axis-gen}) is derived by considering the fact that $\bm{k}^{T}\bm{i} =0$, and $\mathbf{T}_{\bm{k}}$ is a $3\times 3$ matrix whose elements are functions of $\bm{k}$, i.e., 
 \begin{align}
 	\mathbf{T}_{\bm{k}}
 	&\defeq \left[\begin{array}{ccc}
 		0& -k_{z} & k_{y}\\
 		k_{z} & 0 & -k_{x} \\
 		-k_{y} & k_{x} & 0
 	\end{array}\right],
 \end{align}
 with $k_{x}$, $k_{y}$, and $k_{z}$ being the elements of ${\bm{k}}$, i.e., ${\bm{k}}
 	= \left[\begin{array}{ccc}
 		k_{x}& k_{y} & k_{z}
 	\end{array}\right]^{T}$. One can verify that $\bm{j}=\left[\begin{array}{ccc}
 		0& 1 & 0
 	\end{array}\right]^{T}$ if we take $\bm{i}=\left[\begin{array}{ccc}
 		1& 0 & 0
 	\end{array}\right]^{T}$ and $\bm{k}=\left[\begin{array}{ccc}
 		0& 0 & 1
 	\end{array}\right]^{T}$.  Before leaving this subsection, let us define the following projection matrix:
 	\begin{align}
 		\mathbf{P}_{\perp,\bm{k}}
 		&\defeq \mathbf{I} - \dfrac{\bm{k}\bm{k}^{T}}{\bm{k}^{T}\bm{k}}, \label{def-Pk}
 	\end{align}
 	which spans the null space of $\bm{k}$. The matrices $\mathbf{T}_{\bm{k}}$ and $\mathbf{P}_{\perp,\bm{k}}$ will be  frequently used in the following sections to determine the orientations of the sensor and source.

\subsection{Definition of the Orientations}

The orientation of the source relative to the sensor is $-\boldgk{\varphi}_{n}$ as shown in Fig. \ref{fig-illust-coreparamter}. According to the underlying principle of  the image model method, the orientation of the virtual sources varies from one to another.  Based on the previously defined coordinate system for the real and virtual sources, the coordinate system of every virtual source is different. For convenience, we denote the $x$-axis, $y$-axis, and the $z$-axis of the $n$th virtual source as $\bm{i}_{n}$, $\bm{j}_{n}$, and $\bm{z}_{n}$, respectively. Note that $\bm{i}_{n}$, $\bm{j}_{n}$, and $\bm{z}_{n}$ all are vectors in the universal coordinate system of length 3. Let us denote the \emph{orientation angles} of the $n$th source as $(\theta_{n}, \phi_{n})$, where $\phi_{n}$ and $\theta_{n}$ are the azimuth and elevation angles, respectively, and $\theta_{n}$ is the angle between the vector $\bm{k}_{n}$ and $-\boldgk{\varphi}_{n}$. The corresponding  \emph{orientation vector} can then be expressed as
\begin{align}
	\boldgk{\gamma}_{n}
	&\defeq \left[ \begin{array}{c}
		\sin \theta_{n} \cos \phi_{n} \\
		\sin \theta_{n} \sin \phi_{n} \\
		\cos \theta_{n}
	\end{array} \right]. \label{def-ort-source}
\end{align}
In case that the source radiation pattern is known {\em a priori}, which is denoted as $\mathcal{B}^{\prime}(f,\theta, \phi)$ under the coordinate system presented in Fig. \ref{fig-front-direction}, the $\mathcal{B}_{n}(f)$ in (\ref{image-moodel-TF}) can be expressed as
\begin{align} 
	\mathcal{B}_{n}(f)
	&= \mathcal{B}^{\prime}(f,\theta_{n}, \phi_{n}). \label{B-n-B0}
\end{align}
Recall that the $(\theta_{n},\phi_{n})$ are the orientation angles of the $n$th source, which is the direction of the sensor under the source coordinate system.

In a similar way, we denote the $x$-, $y$-, and $z$-axes of the sensor as  $\bm{i}_{\mathrm{mic}}$, $\bm{j}_{\mathrm{mic}}$, and $\bm{z}_{\mathrm{mic}}$, respectively. The orientation of the sensor is defined as the direction of the $n$th source, i.e., $\boldgk{\varphi}_{n}$ in Fig. \ref{fig-illust-coreparamter}, in the sensor coordinate system. Similarly, we define the orientation angle of the microphone sensor as $(\theta_{\mathrm{mic},n}, \phi_{\mathrm{mic},n})$. The corresponding orientation vector is then
\begin{align}
	\boldgk{\gamma}_{\mathrm{mic}, n}
	&\defeq \left[ \begin{array}{c}
		\sin \theta_{\mathrm{mic}, n} \cos \phi_{\mathrm{mic}, n} \\
		\sin \theta_{\mathrm{mic}, n} \sin \phi_{\mathrm{mic}, n} \\
		\cos \theta_{\mathrm{mic}, n}
	\end{array} \right]. \label{gamma-n-mic}
\end{align}
 If we know the directivity pattern of the sensor, which is denoted as $\mathcal{A}^{\prime}(f,\theta, \phi)$, the sensor response $\mathcal{A}_{n}(f)$ in (\ref{image-moodel-TF}) can be expressed as
\begin{align} \label{cal-An}
	\mathcal{A}_{n}(f)
	&= \mathcal{A}^{\prime}(f,\theta_{n,\mathrm{mic}}, \phi_{n,\mathrm{mic}}).
\end{align}
Notice that the orientation of the sensor varies with the sources even though its coordinate system is fixed, which is due to the definition of the orientation.

According to (\ref{image-moodel-TF}), it is clear that the orientations of the sources and sensor given in (\ref{cal-An}) and (\ref{B-n-B0}) are the keys to generalize the classical image model method.

\subsection{Orientations of the Source and  Virtual Sources}

\begin{figure}[t]
	\centering
	\includegraphics{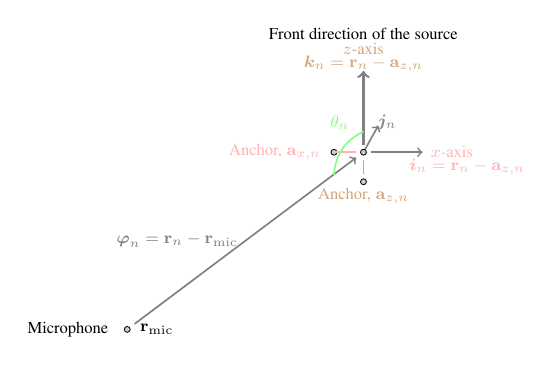}
	\caption{Illustration of the anchor points of the source, where the positions and orientations of the source and sensor are given in Fig.~\ref{fig-illust-coreparamter}}.
	\label{fig-anchor-points}
\end{figure}

To determine the orientations of the sources and sensor, their coordinate system should be built first. In this paper, we introduce anchor points to help building the coordinate system for the $n$th source. As shown in Fig. \ref{fig-anchor-points}, two anchor points are introduced, i.e., $\mathbf{a}_{z,n}$ and $\mathbf{a}_{x,n}$, which correspond to the $z$- and $x$-axes, respectively. For convenience, we denote the anchor points for the real source as
\begin{align}
	\mathbf{a}_{z}
	&\defeq   \left[\begin{array}{ccc}
		x_{\mathrm{s},az}& y_{\mathrm{s},az}& z_{\mathrm{s},az}
	\end{array} \right]^{T}, \label{def-a-z}\\
	\mathbf{a}_{x}
	&\defeq   \left[\begin{array}{ccc}
		x_{\mathrm{s},ax}& y_{\mathrm{s},ax}& z_{\mathrm{s},ax}
	\end{array} \right]^{T}.
\end{align}
In room acoustic simulation, both $\mathbf{a}_{z}$ and $\mathbf{a}_{x}$ should be known or given as the {\em a priori} information. Following the image model method, the images of the anchor points are as follows:
\begin{align}
	\mathbf{a}_{z, n}
	&= \left[\begin{array}{c}
		(-1)^{p_{x}} x_{\mathrm{s}, az} + 2 q_{x} L_{x}\\
		(-1)^{p_{y}} y_{\mathrm{s}, az} + 2 q_{y} L_{y}\\
		(-1)^{p_{z}} z_{\mathrm{s}, az} + 2 q_{z} L_{z}
	\end{array}\right], \label{def-az-n}\\
	\mathbf{a}_{x, n}
	&= \left[\begin{array}{c}
		(-1)^{p_{x}} x_{\mathrm{s}, ax} + 2 q_{x} L_{x}\\
		(-1)^{p_{y}} y_{\mathrm{s}, ax} + 2 q_{y} L_{y}\\
		(-1)^{p_{z}} z_{\mathrm{s}, ax} + 2 q_{z} L_{z}
	\end{array}\right],
\end{align}
where the integers $p_{x}, p_{y}, p_{z}, q_{x}, q_{y},$ and $q_{z}$ are the indices of the $n$th virtual source.

Once the two anchor points of very virtual source are generated, one can determine its $x$- and $z$-axes according to
\begin{align}
	\bm{i}_{n}
	&= \mathbf{r}_{n} - \mathbf{a}_{x, n},\\
	\bm{k}_{n}
	&= \mathbf{r}_{n} - \mathbf{a}_{z, n}.
\end{align}
It follows from (\ref{y-axis-gen}) that the $y$-axis is
\begin{align}
	\bm{j}_{n}
	&= \mathbf{T}_{\bm{k}_{n}}\bm{i}_{n}.
\end{align}
We can further derive that
\begin{align}
	\cos \theta_{n}
	&= - \dfrac{\bm{\varphi}_{n}^{T}\bm{k}_{n}}{\left\|\bm{\varphi}_{n}\right\|\cdot \left\|\bm{k}_{n}\right\|},\label{cos-theta-n}\\
	\sin\theta_{n}
	&= \sqrt{1-\cos^2\theta_{n}},\label{sin-theta-n}\\
	\cos\phi_{n}
	&= - \dfrac{\bm{i}_{n}^{T} \mathbf{P}_{\perp,\bm{k}_{n}} \bm{\varphi}_{n}}{\sqrt{\bm{\varphi}_{n}^{T} \mathbf{P}_{\perp,\bm{k}_{n}} \bm{\varphi}_{n}} \cdot \sqrt{\bm{i}_{n}^{T}\bm{i}_{n}}},\label{cos-phi-n}\\
	\sin\phi_{n}
	&= - \dfrac{\bm{j}_{n}^{T}  \mathbf{P}_{\perp,\bm{k}_{n}}\bm{\varphi}_{n}}{\sqrt{\bm{\varphi}_{n}^{T} \mathbf{P}_{\perp,\bm{k}_{n}}\bm{\varphi}_{n}} \cdot \sqrt{\bm{j}_{n}^{T}\bm{j}_{n}}}.\label{sin-phi-n}
\end{align}
The detailed derivation is shown in Appendix \ref{apdx-orientation}.
Substituting (\ref{cos-theta-n}), (\ref{sin-theta-n}), (\ref{cos-phi-n}), and (\ref{sin-phi-n}) into (\ref{def-ort-source}) gives the orientation vector of the $n$th image source, from which one can determine the source pattern $\mathcal{B}_{n}(f)$ according to (\ref{B-n-B0}).

\subsection{Orientation of the Sensor}
Following the same principle, one can define the two anchor points for the sensor as $\mathbf{a}_{z, \mathrm{mic}}$ and $\mathbf{a}_{x, \mathrm{mic}}$. The $z$-axis, $x$-axis, and $y$-axis can then be expressed as
\begin{align}
	\bm{i}_{\mathrm{mic}}
	&= \mathbf{r}_{\mathrm{mic}} - \mathbf{a}_{x, \mathrm{mic}},\\
	\bm{k}_{\mathrm{mic}}
	&= \mathbf{r}_{\mathrm{mic}} - \mathbf{a}_{z, \mathrm{mic}},\\
	\bm{j}_{\mathrm{mic}}
	&= \mathbf{T}_{\bm{k}_{\mathrm{mic}}}\bm{i}_{\mathrm{mic}}.
\end{align}
It follows immediately that
\begin{align}
	\cos \theta_{\mathrm{mic},n}
	&=  \dfrac{\bm{\varphi}_{n}^{T}\bm{k}_{\mathrm{mic}}}{\left\|\bm{\varphi}_{n}\right\|\cdot \left\|\bm{k}_{\mathrm{mic}}\right\|},\label{cos-theta-n-mic}\\
	\sin\theta_{\mathrm{mic},n}
	&= \sqrt{1-\cos^2\theta_{\mathrm{mic},n}},\label{sin-theta-n-mic}\\
	\cos\phi_{\mathrm{mic},n}
	&=  \dfrac{\bm{i}_{\mathrm{mic}}^{T} \mathbf{P}_{\perp,\bm{k}_{\mathrm{mic}}}\bm{\varphi}_{n}}{\sqrt{\bm{\varphi}_{n}^{T}\mathbf{P}_{\perp,\bm{k}_{\mathrm{mic}}} \bm{\varphi}_{n}} \cdot \sqrt{\bm{i}_{\mathrm{mic}}^{T}\bm{i}_{\mathrm{mic}}}},\label{cos-phi-n-mic}\\
	\sin\phi_{\mathrm{mic},n}
	&=  \dfrac{\bm{j}_{\mathrm{mic}}^{T} \mathbf{P}_{\perp,\bm{k}_{\mathrm{mic}}}\bm{\varphi}_{n}}{\sqrt{\bm{\varphi}_{n}^{T}\mathbf{P}_{\perp,\bm{k}_{\mathrm{mic}}} \bm{\varphi}_{n}} \cdot \sqrt{\bm{j}_{\mathrm{mic}}^{T}\bm{j}_{\mathrm{mic}}}}.
	\label{sin-phi-n-mic}
\end{align}
Substituting (\ref{cos-theta-n-mic}), (\ref{sin-theta-n-mic}), (\ref{cos-phi-n-mic}), and (\ref{sin-phi-n-mic}) into (\ref{gamma-n-mic}) gives the orientation vector of the sensor relative to the $n$th virtual source. One can then obtain the pattern $\mathcal{A}_{n}(f)$ required for the impulse response simulation according to (\ref{cal-An}).

\section{A Special Case: the Radiation Pattern of the Source is Symmetric with Respect to the Front Direction}
\label{sect-specialcase}

\begin{figure}
	\centering
	\includegraphics[width=7.5cm]{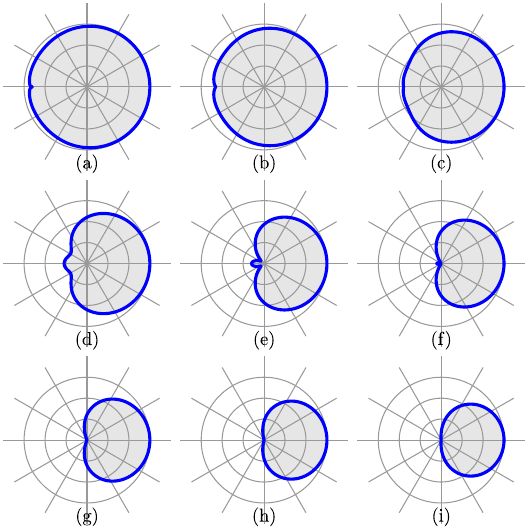}\\
	\caption{Source radiation patterns at different frequencies: (a) $f=250$~Hz, (b) $f=500$~Hz, (c) $f=1$~kHz, (d) $f=2$~kHz, (e) $f=3$~kHz, (f) $f=4$~kHz, (g) $f=5$~kHz, (h) $f=6$~kHz, and (i) $f=7$~kHz.}
	\label{fig-radiation-pattern}
\end{figure}

To help understand the presented method, let us consider the particular case where the radiation pattern of the source is symmetric with respect to its front direction, i.e., $z$-axis, and the directivity pattern of the sensor is omnidirectional, i.e., $\mathcal{A}_{n}(f)=1, \forall f, n$. In this case, only the   $z$-axis direction is necessary to determine the response of the virtual source relative to the sensor. As a result, we need only one anchor point, which is the $\mathbf{a}_{z}$ in (\ref{def-a-z}). To further simplify the implementation of the impulse response simulation, we propose the following simplified source radiation pattern, which is effective to model radiation pattern of a human speaker:
\begin{align}
	 \mathcal{B}_n \left( f \right)
	& =  \varepsilon_{n} \left( f \right) \left[1-\mathcal{S}_{n} \left( f \right) \right] + \mathcal{S}_{n} \left( f \right), \label{B-n-simplify}
\end{align}
where
\begin{align}
	\varepsilon_{n} \left( f \right)
	&\defeq  \dfrac{1}{(1+f_{\mathrm{kHz}})^2}\left[0.5(1-\cos\theta_{n})\right]^{8}, \\
	\mathcal{S}_{n} \left( f \right)
	&\defeq  \left[0.5(1+\cos\theta_{n})\right]^{\rho \left( f \right)}, \label{pattern-cardioid-src}\\
	\rho \left( f \right)
	&\defeq  \ln \left( 1+ 0.6743 f_{\mathrm{kHz}}+
	0.3776 f_{\mathrm{kHz}}^{2}\right. \nonumber\\
	&~~~~ \left. -
	0.0540 f_{\mathrm{kHz}}^{3} +
	0.020 f_{\mathrm{kHz}}^{4} \right),\\
	f_{\mathrm{kHz}}
	&\defeq  f/1000,
\end{align}
with $f_{\mathrm{kHz}}$ being the frequency in kilohertz (kHz) and  $\theta_{n}$ being the angle of the sensor relative to the front direction of the source. Recall that the term $0.5(1+\cos\theta_{n})$ is a cardioid beampattern, so $\mathcal{S}_{n} \left( f \right)$ is actually a cardioid beampattern of order $\rho \left( f \right)$, which is a function of the frequency. The orders of the cardioid beampattern at different frequencies are determined by the radiation pattern of human speakers presented in \cite{flanagan1960analog,direct2020}. Figure~\ref{fig-radiation-pattern} plots the radiation patterns at several frequencies. As seen, the source radiation pattern is almost omnidirectional at low frequencies, and becomes more and more directional as the frequency increases.

\subsection{Source Orientation}

Recall that the source position is $\mathbf{r}_{\mathrm{s}}$, and the $z$-axis of the anchor point is $\mathbf{a}_{z}$. Let us denote the $z$-axis of the real source as
\begin{align}
	\bm{k}^{\prime}_{0}
	\defeq  \mathbf{r}_{\mathrm{s}} - \mathbf{a}_{z},
\end{align}
which is also the front direction of the real source. By considering (\ref{def-r-pq}), (\ref{def-rn}), and (\ref{def-az-n}), the front direction of the $n$th virtual source can be derived as
 \begin{align}
	\bm{k}_{n}
	&=  \mathbf{r}_{n} - \mathbf{a}_{n}
	= \left[\begin{array}{c}
		(-1)^{p_{x}}  \\
		(-1)^{p_{y}}  \\
		(-1)^{p_{z}}
	\end{array}\right] \odot \bm{k}^{\prime}_{0},
\end{align}
where $\odot$ denotes the Hadamard product. Considering that $p_{x}\in\{0,1\}, p_{y}\in\{0,1\}$, and $p_{z}\in\{0,1\}$, one can check that the virtual sources can only have $8$ possible different front directions, which coincides with the results presented in \cite{brinkmann2019extending}. The orientation angle $\theta_{n}$ follows that
\begin{align}
	\cos \theta_{n}
	&= -\dfrac{\bm{\varphi}_{n}^{T}\bm{k}_{n}}{\left\|\bm{\varphi}_{n}\right\|\cdot \left\|\bm{k}_{n}\right\|},
	\label{cos-theta-n-simply}
\end{align}
where $\bm{\varphi}_{n}$ is defined in (\ref{def-varphi-n}). Since the simplified radiation pattern is a function of $\cos\theta_{n}$ as shown in (\ref{B-n-simplify}),  the virtual source pattern $\mathcal{B}_{n} \left( f \right)$ can be obtained by substituting (\ref{cos-theta-n-simply}) into (\ref{pattern-cardioid-src}). Note that the positions of the virtual sources are distributed around the microphone sensor; the value of orientation angle $\theta_{n}$ can vary from $0$ to $\pi$.

\section{Implementation of the Algorithm}
\label{sect-implement}

\subsection{Mathematical Model of the Radiation Patterns}
Generally, the radiation pattern of the source and the directivity pattern of the sensor should be functions of the frequency $f$, as well as the angles $\theta$ and $\phi$, which can be modeled in the form of spherical harmonics. Specifically, the function of the patterns can be expressed as
\begin{align}
	\mathcal{G}(f; \theta, \phi)
	&= \sum_{m=0}^{M}\sum_{\ell=-m}^{m}
	g_{m,\ell}(f) \mathcal{Y}_{m,\ell}(\theta,\phi), \label{G-Y-theta-phi}
\end{align}
where $M$ is the order of the spherical harmonics, $\mathcal{Y}_{m,\ell}(\theta,\phi)$ is the spherical harmonic of order $m$ and degree $\ell$. It is known that the spherical harmonic function can be further expressed as
\begin{align}
	\mathcal{Y}_{m,\ell}(\theta, \phi)
	&= \sqrt{\dfrac{2m+1}{4\pi} \dfrac{(m-\ell)!}{(m+\ell)!}}\mathcal{P}_{m,\ell}(\cos\theta)e^{\displaystyle \jmath \ell \phi}, \label{Y-theta-phi}
\end{align}
where $\mathcal{P}_{m,\ell}(\cos\theta)$ is the associated Legendre polynomial with respect to $\cos\theta$ of order $m$ and degree $\ell$. The spherical harmonics follows the orthogonal property, i.e.,
 \begin{align}
 & \int_{0}^{2\pi}\int_{0}^{\pi} \mathcal{Y}_{m,\ell}(\theta, \phi)\mathcal{Y}_{m^{\prime},\ell^{\prime}}^{\ast}(\theta, \phi)\sin\theta d \theta d \phi \nonumber \\
 &= \delta(m-m^{\prime}) \delta(\ell-\ell^{\prime}).
 \end{align}
 Following the above general model, one can express the sensor directivity and the source radiation patterns as
\begin{align}
	\mathcal{A}^{\prime}(f; \theta, \phi)
	&= \sum_{m=0}^{M}\sum_{\ell=-m}^{m}
	g_{m,\ell}^{a}(f) \mathcal{Y}_{m,\ell}(\theta, \phi),\\
	\mathcal{B}^{\prime}(f; \theta, \phi)
	&= \sum_{m=0}^{M}\sum_{\ell=-m}^{m}
	g_{m,\ell}^{b}(f) \mathcal{Y}_{m,\ell}(\theta, \phi),
\label{B-Y-theta-phi}
\end{align}
where $g^{a}_{m,\ell}(f)$ and $g^{b}_{m,\ell}(f)$ are pattern coefficients, which share the same physical meaning as the $g_{m,\ell}(f)$, and are independent of the angle $\theta$ and $\phi$.

\begin{figure}
	\centering
	\includegraphics{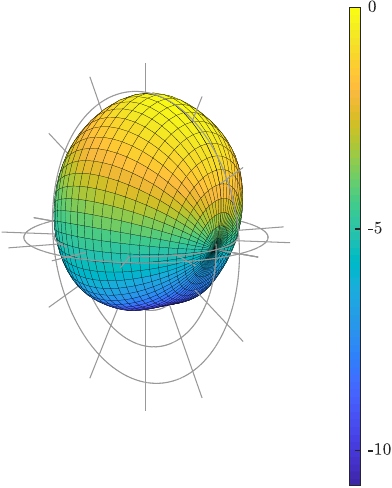}
	\caption{The radiation pattern of a singing speaker at $4$ kHz frequency interpolated by spherical harmonics, where the front direction is along the $z$-axis, the $xz$ and $xy$ planes are marked with the circles and the spikes, and the order of the spherical harmonics is $M=9$.}
	\label{fig-example-3Dpattern}
\end{figure}

Measuring the radiation patterns of sound sources has been intensively studied over the past few decades and several patterns were developed to model commonly seen source sources \cite{flanagan1960analog,direct2020,weinreich1980method,bilbao2019incorporating,ahrens2020interpolation}. For example, the 3D radiation pattern of a singing speaker is shown in Fig. \ref{fig-example-3Dpattern}, where the $xz$  and the $xy$ planes are marked with the circles and the spikes, the front direction of the source is at the $z$-axis, and  the order of the spherical harmonics is $M=9$ (the original data are from \cite{ahrens2020interpolation}). From this figure, one can clearly see that the radiation pattern of a real human speaker is not omnidirectional, and the difference between the front and the back direction responses is more than $10$~dB, which is significant and has to be considered in the room acoustic simulation. Note that measuring the radiation patterns is beyond the main thrust of this work. So, we assume that the pattern coefficients $g_{m,l}(f)$ are known {\em a priori} so the focal point of our work is placed on how to generate the room impulse responses.

There are different ways to calculate the radiation pattern according the given {\em a priori} information. Two straightforward ways are the following.
\begin{itemize}
\item The coefficients $g_{m,\ell}$ are given. In this case, one can calculate the radiation pattern of the source according to $\cos\theta_{n}$ and $e^{\displaystyle\jmath \phi_{n}}$. Note that $\cos\theta_{n}$ can be determined by (\ref{cos-theta-n}), and that  $e^{\displaystyle\jmath \phi_{n}}=\cos\phi_{n} + \jmath \sin \phi_{n}$, with $\cos\phi_{n} $ and $\sin\phi_{n} $ being given in (\ref{cos-phi-n}) and (\ref{sin-phi-n}).

\item Some sample points of $\mathcal{B}^{\prime}(f,\theta, \phi)$ are given, i.e.,
\begin{align}
	\mathcal{B}^{\prime}(f,\theta_{i}, \phi_{i}), i=0,1,2,\ldots,L_{b}-1.
\end{align}
In this case, the $\mathcal{B}_{n}(f)$ can be determined according to
\begin{align}
	\mathcal{B}_{n}(f)
	&=\mathcal{B}^{\prime}(f,\theta_{i^{\ast}}, \phi_{i^{\ast}}),\\
	i^{\ast}
	&= \arg \max_{i\in\{0,1,\ldots,L_{b}-1\}} \boldgk{\gamma}_{i}^{T}\boldgk{\gamma}_{n}.
\end{align}
\end{itemize}
Recall that the orientation  vector $\boldgk{\gamma}_{n} $ is a function of the orientation angles $\theta_{n}$ and $\phi_{n}$, see (\ref{def-ort-source}). The unit vector $\boldgk{\gamma}_{i} $ is a function of $\theta_{i}$ and $\phi_{i}$, which is defined in the same way as $\boldgk{\gamma}_{n} $. As to the sensor directivity pattern, it can be modeled and calculated in the same way as the source radiation patterns.

\subsection{The Fractional Time Delay and the Radiation Pattern}

In practical applications, sub-sample delays, which are fractions of the sampling period, are needed to generate impulse responses that can accurately model the reflection paths. To achieve such delays, let us rewrite the transfer function from the source to the sensor positions in (\ref{image-moodel-TF}) as
\begin{align}
	\ddot{H}(\omega)
	&= \sum_{n=0}^{N-1}
	\dfrac{1}{4\pi d_{n}} \beta_{n} \ddot{\mathcal{C}}_{n}(\omega)
	\expd{-\jmath \omega \left( \ddot{\tau}_{n} + {\zeta}_{n}\right)}\\
	&=\sum_{n=0}^{N-1} \dfrac{1}{4\pi d_{n}} \beta_{n} \left[\ddot{\mathcal{C}}_{n}(\omega)\expd{-\jmath \omega {\zeta}_{n}}\right]
	\expd{-\jmath \omega  \ddot{\tau}_{n} },
\end{align}
where
\begin{align}
	\ddot{\mathcal{C}}_{n} \left( \omega \right)
	&\defeq  {\mathcal{A}}_{n} \left(  \frac{\omega}{2\pi} f_{\mathrm{s}} \right) {\mathcal{B}}_{n} \left(  \frac{\omega}{2\pi} f_{\mathrm{s}} \right),\\
	\ddot{\tau}_{n}
	&\defeq  \lfloor\tau_{n} f_{\mathrm{s}} \rceil,\\
	\zeta_{n}
	& \defeq  \tau_{n} f_{\mathrm{s}} - \ddot{\tau}_{n},
\end{align}
with $\omega \in[0, 2\pi)$ being the angular frequency and $\lfloor\cdot\rceil$ being the rounding operation, e.g., $\lfloor 0.7\rceil = 1$ and $\lfloor 0.3\rceil = 0$.
Note that $\ddot{\tau}_{n}$ is an integer and $\zeta_{n}\in[-0.5,0.5)$ denotes a fractional delay.

To consider the radiation pattern and the fractional delay jointly, let us define a vector of length $2D+1$, i.e.,
\begin{align}
  \ddot{\mathbf{c}}_{n}\defeq \left[\begin{array}{cccc}
		c_{n}(0) &c_{n}(1)& \cdots &c_{n}(2D)
	\end{array} \right]^{T},
\end{align}
where
\begin{align}
	c_{n}(\ell)
	&= e_{n}(D_{e} - D +\ell ),\label{b-n-l}\\
	e_{n}(\ell)
	&\defeq  \dfrac{1}{2\pi}\int_{0}^{2\pi} \ddot{\mathcal{C}}_{n}(\omega)\expd{-\jmath \omega ({\zeta}_{n}+D_{e})} \expd{\jmath \omega \ell} d\omega,
	\label{e-n}
\end{align}
and the parameter $D_{e}$ is introduced in (\ref{e-n}) to ensure that the maximum amplitude of $e_{n}(\ell)$ appears in the middle of the sequence. Generally,
$D_{e}$ and $D$ should satisfy $D_{e}\geq D$. It is clear that $e_{n}(\ell)$ in (\ref{e-n}) can be computed efficiently using the inverse fast Fourier transform (FFT).
One can check that $c_{n}(\ell)=e_{n}(\ell)$ if $D_{e}=D$. In the particular case where $\ddot{\mathcal{C}}_{n}(\omega)=1, \forall \omega$, i.e., the radiation pattern is omnidirectional, we have
\begin{align}
   c_{n}^{\mathrm{omni}}(\ell)
	=\dfrac{\sin\left[(\ell-\zeta_{n}-D)\pi\right]}{(\ell-\zeta_{n}-D)\pi}.
\label{b-omni-expl}
\end{align}
In this case and if $\zeta_{n}=0$, $\ddot{\mathbf{c}}_{n}$ degenerates to a one vector of length $2D+1$.

Finally, to reduce aliasing, an anti-aliasing window (e.g., the Hamming window)  of length $2D+1$ should be applied to the $\ddot{\mathbf{c}}_{n}$ vector. Let us denote the window as $\boldgk{\psi}_{n}\defeq \left[\begin{array}{cccc}
		\psi_{n}(0) &\psi_{n}(1)& \cdots &\psi_{n}(2D)
	\end{array} \right]^{T}$. If the Hamming window is used, its elements are then as follows:
	\begin{align}
		\psi_{n}(\ell )
		&= 0.54 - 0.46 \cos\left[\dfrac{\pi (\ell-\zeta_{n})}{D}\right].\label{cal-window}
	\end{align}
More detailed discussion about how to deal with fractional time delay can be found in \cite{Laakso1996}.

\subsection{Implementation}
\label{sect-alg-implemt}

In this subsection, we discuss how to implement the presented algorithm according to the derivations given in the previous subsections. Through analysis and simulations, it is seen that the low-order images generally follow source radiation pattern while the high-order images (corresponding to the paths from the source to the sensor positions via multiple reflections) become more or less omnidirectional. Therefore, if the computational cost is a concern, one can reduce the cost by treating the higher-order images as omnidirectional virtual sources. Let us define an integer $Q_{\max}$. The image is considered to be omnidirectional if $|q_{x}|> Q_{\max}$,  $|q_{y}|> Q_{\max}$, or  $|q_{z}|> Q_{\max}$. The total number of the images with directional radiation patterns is then $8(2 Q_{\max}+1)^3$. It can be noticed that, by taking $Q_{\max}<0$, the proposed algorithm degenerates to the traditional image model method.

Let us define the impulse response vector of length $L_{h}$ as
	$\mathbf{h}
	\defeq  \left[\begin{array}{cccc}
		h(0) &h(1)& \cdots &h(L_{h}-1)
	\end{array} \right]^{T}.$
The impulse response is then generated by the following three steps.
\begin{itemize}
\item[1)] Specify the parameters $Q_{x}$, $Q_{y}$, $Q_{z}$, $D$, the sound speed $c$, the reflection coefficients  $\beta_{x_{0}},\beta_{x_{1}}, \beta_{y_{0}}, \beta_{y_{1}}, \beta_{z_{0}}, \beta_{z_{0}}$, the source position $\mathbf{r}_{\mathrm{s}}$, the microphone position $\mathbf{r}_{\mathrm{mic}}$, and the anchor points.
\item[2)] For every value of $n$, or equivalently every $(p_{x}, p_{y}, p_{z}; q_{x}, q_{y}, q_{z})$,  compute $d_{n}$,  $\ddot{\tau}_{n}$, $\zeta_{n}$, $\ddot{\mathbf{c}}_{n}$, $\boldgk{\psi}_{n}$, and $\beta_{n}$.
\item[3)] If the images are viewed as lower-order virtual sources, compute $\ddot{\mathbf{b}}_{n}$ according to (\ref{b-n-l}) and (\ref{e-n})  and then update the impulse response according to
	\begin{align}
		\left[\mathbf{h}\right]_{\ddot{\tau}_{n}-D:\ddot{\tau}_{n}+D}
		&\leftarrow   \left[\mathbf{h}\right]_{\ddot{\tau}_{n}-D:\ddot{\tau}_{n}+D}
		+ \dfrac{ \beta_{n}}{4\pi d_{n}} \boldgk{\psi}_{n}\odot \ddot{\mathbf{c}}_{n} . \label{update-h}
	\end{align}
	Otherwise, we treat the images and the sensor as omnidirectional and update the impulse response according to
	\begin{align}
		\left[\mathbf{h}\right]_{\ddot{\tau}_{n}-D:\ddot{\tau}_{n}+D}
		&\leftarrow   \left[\mathbf{h}\right]_{\ddot{\tau}_{n}-D:\ddot{\tau}_{n}+D}
		+ \dfrac{\beta_{n}}{4\pi d_{n}}  \boldgk{\psi}_{n}\odot \ddot{\mathbf{c}}^{\mathrm{omni}}_{n}. \label{update-h-omni}
	\end{align}
Recall that the element of $\ddot{\mathbf{c}}^{\mathrm{omni}}_{n}$ has an explicit form [see (\ref{b-omni-expl})]. So, updating (\ref{update-h-omni}) is computationally very efficient.
\end{itemize}
Finally, the detailed steps for implementing the proposed algorithm is given in Appendix \ref{apdx-summarize-imagemethod}.

\begin{figure}[t!]
	\centering
	\includegraphics[width=7.5cm]{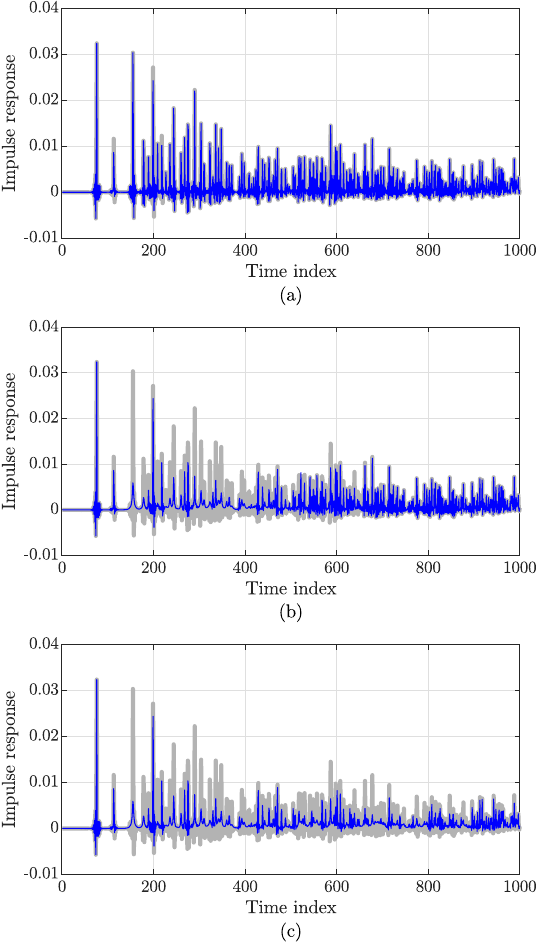}
	\caption{Generated impulse responses for a directional source with different values of $Q_{\max}$: (a) $Q_{\max}=0$, (b) $Q_{\max}=1$, and (c)  $Q_{\max}=2$. The gray line shows the impulse response for the omnidirectional source under the same conditions.}
	\label{fig-impulseResponse}
\end{figure}

\section{Examples of the Generated Impulse Responses}
\label{sect-examples}

\begin{figure*}
	\centering
	\includegraphics{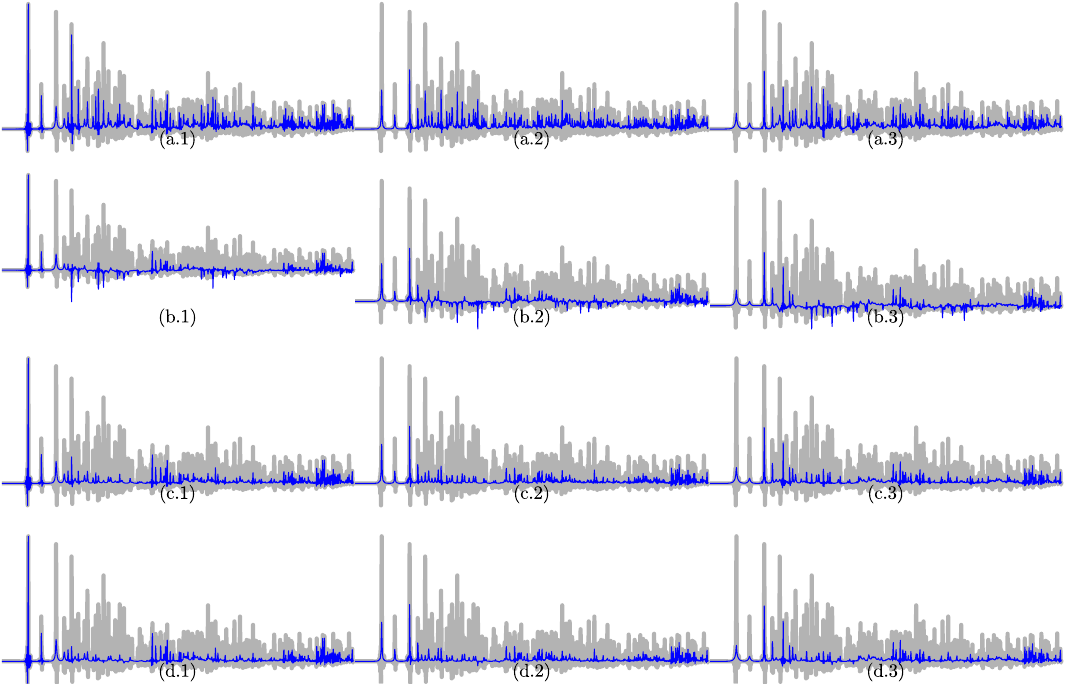}
	\caption{The impulse responses generated by the proposed method under different sensor types and source orientations with an ideal source pattern, where the sensor directivity pattern and the source orientation condition of every subfigure is listed in Tab.~\ref{tab-condit}. }
	\label{fig-irs-DsDm}
\end{figure*}

\begin{figure*}
	\centering
	\includegraphics{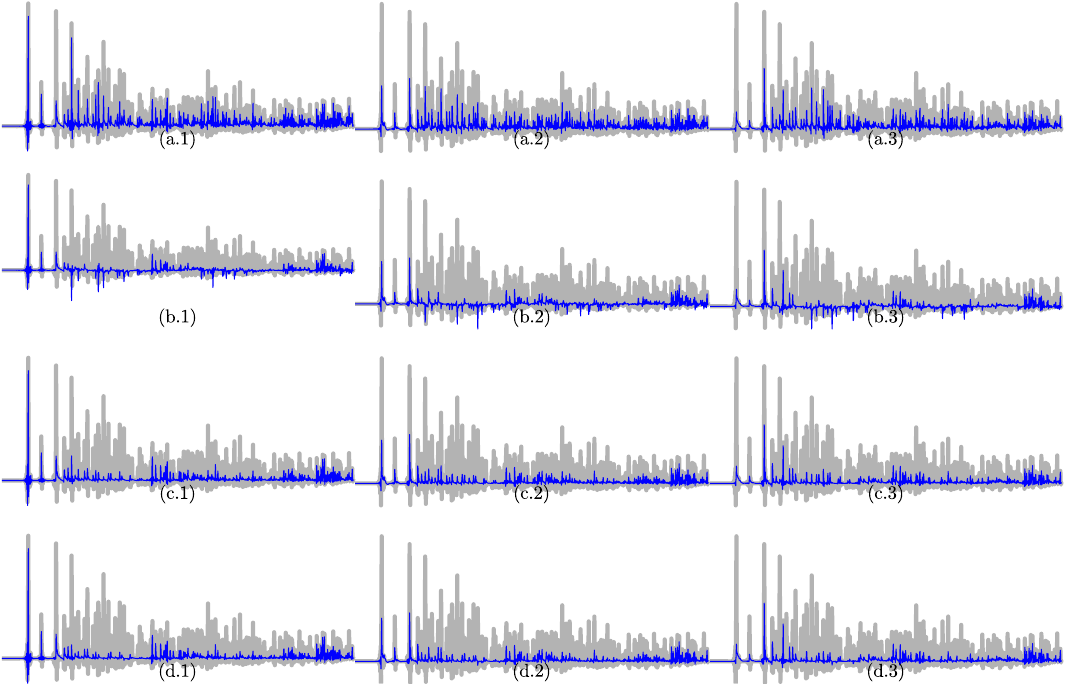}
	\caption{The impulse responses generated by the proposed method under different sensor types and source orientations with a measured source pattern, where the sensor directivity pattern and the source orientation condition of every subfigure is listed in Tab.~\ref{tab-condit}. }
	\label{fig-irs-DsDm-sMd}
\end{figure*}

\begin{table}
\centering
\caption{Conditions for the Subfigures in Figs~\ref{fig-irs-DsDm} and \ref{fig-irs-DsDm-sMd}. }
\begin{tabular}{l|lll} \hline
                & $0^{\circ}$ deviation     & $90^{\circ}$ deviation   & $180^{\circ}$ deviation    \\ \hline
Omnidirectional & (a.1) & (a.2) & (a.3)  \\
Dipole          & (b.1) & (b.2) & (b.3)  \\
Cardioid        & (c.1) & (c.2) & (c.3)  \\
Supercardioid   & (d.1) & (d.2) & (d.3) \\ \hline
\end{tabular} \vskip 10pt
\label{tab-condit}
\end{table}

Since the image model method has been widely adopted in various fields, we will use this method as the baseline and show how the directivity and the radiation patterns can affect the impulse responses. Two types of the source radiation patterns are considered: one is the simplified source radiation pattern presented in (\ref{B-n-simplify}), and the other is the singing voice radiation pattern presented in \cite{ahrens2020interpolation}. Other simulation conditions are as follows. The sampling rate is $16$~kHz. The speed of sound is $c=340$~m/s. The length of the impulse response is $L_{h}=2048$. The room size is $L_{x}=4$~m, $L_{y}=4$~m, and $L_{z}=4$~m. The reflection coefficients are ($0.96, 0.8, 0.96, 0.9, 0.5, 0.5$). The position of the source is at $\mathbf{r}_{\mathrm{s}} = (3, 3, 1)$, and the microphone is at $\mathbf{r}_{\mathrm{mic}} = (1.5, 1.5, 1)$. The relative positions of the source and microphone are as illustrated in Fig.~\ref{fig-illust-coreparamter}.

To visualize the influence of the source orientation, three source orientation conditions are considered in the simulations.
\begin{itemize}
	\item The front direction of the source faces the sensor direction. In this case, the anchors of the $z$-and $x$-axes are $\mathbf{a}_{z}=(3.1,3.1,1)$ and $\mathbf{a}_{x}=(2.9,3.1,1)$, respectively.
	\item The front direction of the source deviates from the sensor direction by $90^{\circ}$. In this case, the anchors are $\mathbf{a}_{z}=(3.1,2.9,1)$ and $\mathbf{a}_{x}=(2.9,2.9,1)$.
	\item The front direction of the source deviates from the sensor direction by $180^{\circ}$. In this case, the anchors are $\mathbf{a}_{z}=(2.9,2.9,1)$ and $\mathbf{a}_{x}=(2.9,3.1,1)$.
\end{itemize}
 Inspired by the gradient microphone \cite{Olson1946} and the differential beamforming \cite{Chen2014,Benesty2016,Pan2015a,Pan2019}, we consider four types of sensor directivity patterns.
\begin{itemize}
	\item \emph{Omnidirectional}. This case corresponds to $\mathcal{A}^{\prime}(f,\theta,\phi) = 1$.
	\item \emph{Dipole}. This case corresponds to $\mathcal{A}^{\prime}(f,\theta,\phi) = \cos\theta$.
	\item \emph{Cardioid}. This case corresponds to $\mathcal{A}^{\prime}(f,\theta,\phi) = 0.5 + 0.5\cos\theta$.
	\item \emph{Supercardioid}. This case corresponds to $\mathcal{A}^{\prime}(f,\theta,\phi) = (\sqrt{2}-1) + (2-\sqrt{2})\cos\theta$.
\end{itemize}
Recall that $\theta$ is the angle between the front direction of the sensor and the front direction of the source. It can be verified that $\mathcal{A}^{\prime}(f,\theta,\phi)=1$ if the source is at the front direction of the sensor, i.e., $\theta=0$. 

According to the discussion in Section \ref{sect-alg-implemt}, the parameter $Q_{\max}$ plays an important role on the complexity of the proposed algorithm. It determines how many virtual sources are considered directional, as well as the corresponding sensor directivity.  Figure~\ref{fig-impulseResponse} plots the impulse responses generated by the developed method with three different values of $Q_{\max}$ (only the first 1000 coefficients of the impulse responses are plotted), where (a) $Q_{\max}=0$, (b) $Q_{\max}=1$, and (c) $Q_{\max}=2$. The gray line shows the impulse response without considering the source radiation pattern and the sensor directivity pattern. As seen, more and more coefficients are attenuated as the $Q_{\max}$ increases. Through our investigation, we found that $Q_{\max}=2$ is enough to cover the strong reflection pathes for most practical applications. Therefore, in the rest simulations, we set $Q_{\max}=2$.

Figure~\ref{fig-irs-DsDm} plots the generated impulse responses with different sensor directivity patterns and source orientations, where the source radiation pattern is in the form of (\ref{B-n-simplify}). The conditions for each subfigure are presented in Tab.~\ref{tab-condit}. As seen, the amplitude of the direct path decreases as the deviation between the source front direction and the sensor direction increases [see Fig.~\ref{fig-irs-DsDm}(a.1), (a.2), and (a.3)]. Comparing the first row of Fig. \ref{fig-irs-DsDm} with the other three rows, one can see that the amplitude of the reflections are further attenuated in the case that the source directivity pattern is considered.  Figure~\ref{fig-irs-DsDm-sMd} plots the generated impulse response where source radiation pattern is the one for a singing speaker \cite{ahrens2020interpolation}). As seen, the results are similar to those in Fig. \ref{fig-irs-DsDm}, which corroborates the radiation pattern model presented in (\ref{B-n-simplify}).

\section{Conclusions}
\label{sect-cnls}

This paper studied the problem of room acoustic simulation, and extended the image model method, which assumes that both the source and sensor are omnidirectional, to the generic case, which includes directional sources and sensors with any pre-specified radiation/directivity  patterns. A model was presented to model the radiation pattern of directional sources, which is a function of the frequency and the source orientation relative to the sensor. The model is particularly useful for simulating the room impulse response when the source is a human speaker. We presented a method to determine the orientation vector and angles for virtual sources through introducing anchor points to the source. By considering the source radiation pattern, the sensor directivity pattern as well as the fractional time delay simultaneously, a method was developed to generate the room impulse responses. We further presented a simplified version of the developed method, in which the lower-order images are assumed to be directional and their radiation/directivity patterns follow, respectively, the pre-specified source radiation and sensor directivity patterns while the higher-order images are assumed to be omnidirectional. This simplification follows the sound propagation principle and can greatly reduce the complexity of the algorithm, thereby making the room acoustic simulation computationally efficient.

\appendix
\section{Derivation of the Source Orientations}
\label{apdx-orientation}

\begin{figure}
	\centering
	\includegraphics{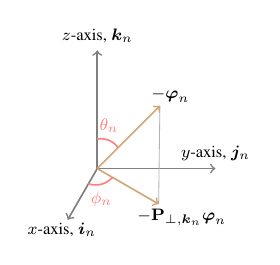}
	\caption{Illustration of the source orientation angles, where $\boldgk{\theta}_{n}$ and $\boldgk{\phi}_{n}$ are the elevation and azimuth angles, respectively.}
	\label{fig-theta-phi}
\end{figure}

Let us first illustrate in Fig.~\ref{fig-theta-phi} the four vectors: $\bm{i}_{n}$, $\bm{j}_{n}$, $\bm{k}_{n}$, and $-\boldgk{\varphi}_{n}$. As seen, the elevation angle $\theta_{n}$ is the angle between the vectors $-\boldgk{\varphi}_{n}$ and $\bm{k}_{n}$. Then, it follows that
\begin{align}
	\cos \theta_{n}
	&= - \dfrac{\bm{\varphi}_{n}^{T}\bm{k}_{n}}{\left\|\bm{\varphi}_{n}\right\|\cdot \left\|\bm{k}_{n}\right\|}.
\end{align}
Due to the fact that $\theta_{n}$ varies only from $0$ to $\pi$, the value of $\sin\theta_{n}$ is nonnegative. We then have
\begin{align}
	\sin\theta_{n}
	&= \sqrt{1-\cos^2\theta_{n}}.
\end{align}
According to the definition of the projection matrix $\mathbf{P}_{\perp,\bm{k}_{n}}$ in (\ref{def-Pk}), the projection of $-\boldgk{\varphi}_{n}$ on the $xy$ plane can be expressed as
\begin{align}
	\boldgk{\vartheta}_{n}
	&= -\mathbf{P}_{\perp,\bm{k}_{n}}\boldgk{\varphi}_{n}.
\end{align}
Since the azimuth angle is defined as the angle between the $\bm{i}_{n}$ and $\boldgk{\vartheta}_{n}$, we have
\begin{align}
	\cos\phi_{n}
	&= \dfrac{\bm{i}_{n}^{T}\boldgk{\vartheta}_{n}}{\sqrt{\boldgk{\vartheta}_{n}^{T}\boldgk{\vartheta}_{n}} \cdot  \sqrt{\bm{i}_{n}^{T}\bm{i}_{n}}}\\
	&= -\dfrac{\bm{i}_{n}^{T}\mathbf{P}_{\perp,\bm{k}_{n}}\boldgk{\varphi}_{n}}{\sqrt{\boldgk{\varphi}_{n}^{T}\mathbf{P}^{T}_{\perp,\bm{k}_{n}}\mathbf{P}_{\perp,\bm{k}_{n}}\boldgk{\varphi}_{n}} \cdot  \sqrt{\bm{i}_{n}^{T}\bm{i}_{n}}}\label{apdx-phin-1}\\
	&= -\dfrac{\bm{i}_{n}^{T}\mathbf{P}_{\perp,\bm{k}_{n}}\boldgk{\varphi}_{n}}{\sqrt{\boldgk{\varphi}_{n}^{T}\mathbf{P}_{\perp,\bm{k}_{n}}\boldgk{\varphi}_{n}} \cdot  \sqrt{\bm{i}_{n}^{T}\bm{i}_{n}}},\label{apdx-phin-2}
\end{align}
where (\ref{apdx-phin-2}) is derived  from (\ref{apdx-phin-1}) by considering that $\mathbf{P}^{T}_{\perp,\bm{k}_{n}}\mathbf{P}_{\perp,\bm{k}_{n}}=\mathbf{P}_{\perp,\bm{k}_{n}}$.
In a similar way, one can deduce that
\begin{align}
	\cos\left(\dfrac{\pi}{2} - \phi_{n}\right)
	&=\sin\phi_{n}\\
	&= -\dfrac{\bm{j}_{n}^{T}\mathbf{P}_{\perp,\bm{k}_{n}}\boldgk{\varphi}_{n}}{\sqrt{\boldgk{\varphi}_{n}^{T}\mathbf{P}_{\perp,\bm{k}_{n}}\boldgk{\varphi}_{n}} \cdot  \sqrt{\bm{j}_{n}^{T}\bm{j}_{n}}}.\label{apdx-phin-sin}
\end{align}

\section{Implementation of the Developed, Generalized Image Model Method}
\label{apdx-summarize-imagemethod}

Implementation of the generalized image model impulse response simulation method is summarized in Tab.~II.
\vskip 10pt
\begin{table}
  \label{tab-IR-generalazation}
  \caption{Algorithm for APIM impulse response simulation.}
\begin{tabular}{l} %
  \hline
  \hline
  {\huge $\cdot$} Inputs:\\
  ~~{--} length of the impulse response $L_{h}$  \\
  ~~{--} Sound speed $c$, sampling rate $f_{\mathrm{s}}$ \\
  ~~{--} Room size: ($L_{x}, L_{y}, L_{z}$)  \\
  ~~{--} Reflection coefficients: ($\beta_{x_{0}}, \beta_{x_{1}}, \beta_{y_{0}}, \beta_{y_{1}}, \beta_{z_{0}}, \beta_{z_{1}}$)  \\
  ~~{--} Microphone position and anchors: $\mathbf{r}_{\mathrm{mic}}, \mathbf{a}_{\mathrm{mic},z}, \mathbf{a}_{\mathrm{mic},x}$\\
  ~~{--} Source position and anchors: $\mathbf{r}_{\mathrm{s}}, \mathbf{a}_{z}, \mathbf{a}_{x}$\\
  ~~{--} Maximum reflection order: $Q_{x}, Q_{y}, Q_{z}$\\
  ~~{--} Maximum directional order: $Q_{\mathrm{max}}$\\
  ~~{--} Half of window length: $D$\\
  {\huge $\cdot$} For $p_{x}, p_{y}, p_{z}, q_{x}, q_{y}, q_{z}$\\
  \hskip 20pt {\large $\cdot$} Calculate  $\mathbf{r}_{n}$ according to (\ref{def-r-pq}) and (\ref{def-rn}) \\
  \hskip 20pt {\large $\cdot$} Calculate  $d_{n}$ according to (\ref{def-dn}) \\
  \hskip 20pt {\large $\cdot$} Calculate  $\beta_{n}$ according to (\ref{def-beta-n}) \\
  \hskip 20pt {\large $\cdot$} Calculate  $\ddot{c}_{n}$ according to Tab.~III \\
  \hskip 20pt {\large $\cdot$} Calculate  $\boldgk{\psi}_{n}$ according to (\ref{cal-window})\\
  \hskip 20pt {\large $\cdot$} Update  $\mathbf{h}$ according to (\ref{update-h}) and (\ref{update-h-omni})\\
    ~~~~end \\
 {\huge $\cdot$} Return the impulse response $\mathbf{h}$
 \\ \hline
\end{tabular}
\end{table}
Note that the computation of $\ddot{\mathbf{c}}_{n}$ depends on the simulation conditions. If both sensor and the source are omnidirectional, this computation is needed; otherwise, $\ddot{\mathbf{c}}_{n}$ should be computed according to the steps presented in Section \ref{sect-alg-implemt}, which is also summarized in Tab.~III.

\vskip 10pt
\begin{table}
 \label{tab-cal-cn}
  \caption{Algorithm for the computation of $\ddot{\mathbf{c}}_{n}$.}
\begin{tabular}{l} %
  \hline
  \hline
  {\large $\cdot$} Inputs:\\
  ~~{--} Microphone position and anchors: $\mathbf{r}_{\mathrm{mic}}, \mathbf{a}_{\mathrm{mic},z}, \mathbf{a}_{\mathrm{mic},x}$\\
  ~~{--} Source position and anchors: $\mathbf{r}_{n}, \mathbf{a}_{z}, \mathbf{a}_{x}$\\
    ~~{--} Half of window length: $D$ and $D_{e}$\\
    ~~{--} Fractional delay: $\zeta_{n}$ \\
    ~~{--} $p_{x}, p_{y}, p_{z}, q_{x}, q_{y}, q_{z}$\\
      {\large $\cdot$} Calculate key vector: $ \boldgk{\varphi}_{n}$\\
    {\large $\cdot$} Calculate source anchors: $ \mathbf{a}_{z, n}, \mathbf{a}_{x, n}$\\
   {\large $\cdot$} Calculate source coordinate system $\bm{i}_{n}, \bm{j}_{n}, \bm{k}_{n}$ \\
   {\large $\cdot$} Calculate source orientation $\boldgk{\gamma}_{n}, \theta_{n}, \phi_{n}$ \\
   {\large $\cdot$} Calculate sensor orientation $\boldgk{\gamma}_{\mathrm{mic},n}, \theta_{\mathrm{mic},n}, \phi_{\mathrm{mic},n}$ \\
   {\large $\cdot$} Calculate $\mathcal{A}_{n}(\omega)$ and $\mathcal{B}_{n}(\omega)$   \\
   {\large $\cdot$} Calculate  $e_{n}(\ell)$, $c_{n}(\ell)$, and $\ddot{\mathbf{c}}_{n}$  \\
 {\large $\cdot$} Return the vector $\ddot{\mathbf{c}}_{n}$
 \\ \hline
\end{tabular}
\end{table}

\bibliographystyle{elsarticle-num}

\begin{thebibliography}{10}
\expandafter\ifx\csname url\endcsname\relax
  \def\url#1{\texttt{#1}}\fi
\expandafter\ifx\csname urlprefix\endcsname\relax\def\urlprefix{URL }\fi
\expandafter\ifx\csname href\endcsname\relax
  \def\href#1#2{#2} \def\path#1{#1}\fi

\bibitem{Allen1979}
J.~B. Allen, D.~A. Berkley, Image method for efficiently simulating small-room
  acoustics, J. Acoust. Soc. Amer. 65~(4) (1979) 943--950.

\bibitem{Savioja2015}
L.~Savioja, U.~P. Svensson, Overview of geometrical room acoustic modeling
  techniques, J. Acoust. Soc. Amer. 138~(2) (2015) 708--730.

\bibitem{Vaelimaeki2012}
V.~V{\"a}lim{\"a}ki, J.~D. Parker, L.~Savioja, J.~O. Smith, J.~S. Abel, Fifty
  years of artificial reverberation, {IEEE/ACM} Trans. Audio, Speech, Lang.
  Process. 20~(5) (2012) 1421--1448.

\bibitem{savioja1999creating}
L.~Savioja, J.~Huopaniemi, T.~Lokki, R.~V{\"a}{\"a}n{\"a}nen, Creating
  interactive virtual acoustic environments, J. Audio Eng. Soc. 47~(9) (1999)
  675--705.

\bibitem{Borish1984}
J.~Borish, Extension of the image model to arbitrary polyhedra, J. Acoust. Soc.
  Amer. 75~(6) (1984) 1827--1836.

\bibitem{Lehmann2008}
E.~A. Lehmann, A.~M. Johansson, Prediction of energy decay in room impulse
  responses simulated with an image-source model, J. Acoust. Soc. Amer. 124~(1)
  (2008) 269--277.

\bibitem{lehmann2009diffuse}
E.~A. Lehmann, A.~M. Johansson, Diffuse reverberation model for efficient
  image-source simulation of room impulse responses, {IEEE/ACM} Trans. Audio,
  Speech, Lang. Process. 18~(6) (2009) 1429--1439.

\bibitem{jarrett2012rigid}
D.~P. Jarrett, E.~A. Habets, M.~R. Thomas, P.~A. Naylor, Rigid sphere room
  impulse response simulation: Algorithm and applications, J. Acoust. Soc.
  Amer. 132~(3) (2012) 1462--1472.

\bibitem{peterson1986simulating}
P.~M. Peterson, Simulating the response of multiple microphones to a single
  acoustic source in a reverberant room, J. Acoust. Soc. Amer. 80~(5) (1986)
  1527--1529.

\bibitem{betlehem2012sound}
T.~Betlehem, M.~Poletti, Sound field of a directional source in a reverberant
  room, J. Acoust. Soc. New Zealand 25~(4) (2012) 12--22.

\bibitem{kompis1993simulating}
M.~Kompis, N.~Dillier, Simulating transfer functions in a reverberant room
  including source directivity and head-shadow effects, J. Acoust. Soc. Amer.
  93~(5) (1993) 2779--2787.

\bibitem{samarasinghe2018spherical}
P.~N. Samarasinghe, T.~D. Abhayapala, Y.~Lu, H.~Chen, G.~Dickins, Spherical
  harmonics based generalized image source method for simulating room
  acoustics, J. Acoust. Soc. Amer. 144~(3) (2018) 1381--1391.

\bibitem{brinkmann2019extending}
F.~Brinkmann, V.~Erbes, S.~Weinzierl, Extending the closed form image source
  model for source directivity, Technische Universit{\"a}t Berlin, 2019.

\bibitem{Huang2006}
Y.~Huang, J.~Benesty, J.~Chen, Acoustic {MIMO} Signal Processing,
  Springer-Verlag, Berlin, Germany, 2006.

\bibitem{flanagan1960analog}
J.~L. Flanagan, Analog measurements of sound radiation from the mouth, J.
  Acoust. Soc. Amer. 32~(12) (1960) 1613--1620.

\bibitem{direct2020}
C.~P{\"o}rschmann, J.~M. Arend, Analyzing the directivity patterns of human
  speakers, in: DAGA, 2020, pp. 16--19.

\bibitem{weinreich1980method}
G.~Weinreich, E.~B. Arnold, Method for measuring acoustic radiation fields, J.
  Acoust. Soc. Amer. 68~(2) (1980) 404--411.

\bibitem{bilbao2019incorporating}
S.~Bilbao, J.~Ahrens, B.~Hamilton, Incorporating source directivity in
  wave-based virtual acoustics: Time-domain models and fitting to measured
  data, J. Acoust. Soc. Amer. 146~(4) (2019) 2692--2703.

\bibitem{ahrens2020interpolation}
J.~Ahrens, S.~Bilbao, Interpolation and range extrapolation of sound source
  directivity based on a spherical wave propagation model, in: Proc. {IEEE}
  Int. Conf. Acoust., Speech, Signal Process. ({ICASSP}), IEEE, 2020, pp.
  4662--4666.

\bibitem{Laakso1996}
T.~I. Laakso, V.~Valimaki, M.~Karjalainen, U.~K. Laine, Splitting the unit
  delay: {FIR} / all-pass filters design, {IEEE} Signal Process. Mag. 13~(1)
  (1996) 30--60.

\bibitem{Olson1946}
H.~F. Olson, Gradient microphones, J. Acoust. Soc. Amer. 17 (1946) 192--198.

\bibitem{Chen2014}
J.~Chen, J.~Benesty, C.~Pan, On the design and implementation of linear
  differential microphone arrays, J. Acoust. Soc. Amer. 136 (2014) 3097--3113.

\bibitem{Benesty2016}
J.~Benesty, J.~Chen, C.~Pan, Fundamentals of Differential Beamforming, Springer
  Briefs in Electrical and Computer Engineering, Singerpore, 2016.

\bibitem{Pan2015a}
C.~Pan, J.~Chen, J.~Benesty, Theoretical analysis of differential microphone
  array beamforming and an improved solution, {IEEE/ACM} Trans. Audio, Speech,
  Lang. Process. 23~(11) (2015) 2093--2105.

\bibitem{Pan2019}
C.~Pan, J.~Chen, J.~Benesty, S.~Guangming, On the design of target beampatterns
  for differential microphone arrays, {IEEE/ACM} Trans. Audio, Speech, Lang.
  Process. 28~(8) (2019) 1295--1307.

\end{thebibliography}







\end{document}